\documentclass[aps,preprintnumbers,prl,twocolumn,superscriptaddress]{revtex4}

\usepackage{epsfig,latexsym,cancel,amssymb,amsmath}
\usepackage{graphicx}
\usepackage{epstopdf}
\usepackage{color}
\usepackage{graphicx}
\usepackage{dcolumn}
\usepackage{bm}
\usepackage{hyperref}
\usepackage{epstopdf}
\usepackage{color}
\usepackage{hyperref}
\usepackage{epsfig,latexsym,cancel,amssymb,amsmath}

\def\be{\begin{eqnarray}}
\def\ee{\end{eqnarray}}
\def\bea{\begin{eqnarray}}
\def\eea{\end{eqnarray}}

\def\bbuildrel#1_#2^#3{\mathrel{\mathop{\kern 0pt#1}\limits_{#2}^{#3}}}
\def\slash#1{\setbox0=\hbox{$#1$}#1\hskip-\wd0\dimen0=5pt\advance
       \dimen0 by-\ht0\advance\dimen0 by\dp0\lower0.5\dimen0\hbox
         to\wd0{\hss\sl/\/\hss}}

\newcommand{\gae}{\lower 2pt \hbox{$\, \buildrel {\scriptstyle >}\over {\scriptstyle
\sim}\,$}}
\newcommand{\lae}{\lower 2pt \hbox{$\, \buildrel {\scriptstyle <}\over {\scriptstyle
\sim}\,$}}

\newcommand{\beq}{\begin{eqnarray}}
\newcommand{\eeq}{\end{eqnarray}}

\newcommand{\ba}{\begin{array}}
\newcommand{\ea}{\end{array}}

\long\def\symbolfootnote[#1]#2{\begingroup%
\def\thefootnote{\fnsymbol{footnote}}\footnote[#1]{#2}\endgroup}

\def\lsim{\mathrel{\rlap{\lower4pt\hbox{\hskip1pt$\sim$}}
    \raise1pt\hbox{$<$}}}         
\def\gsim{\mathrel{\rlap{\lower4pt\hbox{\hskip1pt$\sim$}}
    \raise1pt\hbox{$>$}}}         

\def\lsim{\:\raisebox{-0.5ex}{$\stackrel{\textstyle<}{\sim}$}\:}
\def\gsim{\:\raisebox{-0.5ex}{$\stackrel{\textstyle>}{\sim}$}\:}

\def\issue(#1,#2,#3){{\bf #1}, #2 (#3)}
\def\opcit(#1){ {\em op. cit.}, #1}

\def\APP(#1,#2,#3){Acta Phys.\ Polon.\ \issue(#1,#2,#3)}
\def\ARNPS(#1,#2,#3){Ann.\ Rev.\ Nucl.\ Part.\ Sci.\ \issue(#1,#2,#3)}
\def\CPC(#1,#2,#3){Comp.\ Phys.\ Comm.\ \issue(#1,#2,#3)}
\def\CIP(#1,#2,#3){Comput.\ Phys.\ \issue(#1,#2,#3)}
\def\EPJC(#1,#2,#3){Eur.\ Phys.\ J.\ C\ \issue(#1,#2,#3)}
\def\EPJD(#1,#2,#3){Eur.\ Phys.\ J. Direct\ C\ \issue(#1,#2,#3)}
\def\IEEETNS(#1,#2,#3){IEEE Trans.\ Nucl.\ Sci.\ \issue(#1,#2,#3)}
\def\IJMP(#1,#2,#3){Int.\ J.\ Mod.\ Phys. \issue(#1,#2,#3)}
\def\JHEP(#1,#2,#3){J.\ High Energy Physics \issue(#1,#2,#3)}
\def\JPG(#1,#2,#3){J.\ Phys.\ G \issue(#1,#2,#3)}
\def\MPL(#1,#2,#3){Mod.\ Phys.\ Lett.\ \issue(#1,#2,#3)}
\def\NP(#1,#2,#3){Nucl.\ Phys.\ \issue(#1,#2,#3)}
\def\NIM(#1,#2,#3){Nucl.\ Instrum.\ Meth.\ \issue(#1,#2,#3)}
\def\PL(#1,#2,#3){Phys.\ Lett.\ \issue(#1,#2,#3)}
\def\PRD(#1,#2,#3){Phys.\ Rev.\ D \issue(#1,#2,#3)}
\def\PRL(#1,#2,#3){Phys.\ Rev.\ Lett.\ \issue(#1,#2,#3)}
\def\SJNP(#1,#2,#3){Sov.\ J. Nucl.\ Phys.\ \issue(#1,#2,#3)}
\def\ZPC(#1,#2,#3){Zeit.\ Phys.\ C \issue(#1,#2,#3)}

\def\beq{\begin{eqnarray}}
\def\eeq{\end{eqnarray}}
\def\bea{\begin{eqnarray}}
\def\eea{\end{eqnarray}}
\def\to{\rightarrow}

\begin{document}
\DeclareGraphicsExtensions{.jpg,.pdf,.mps,.png,}

\preprint{PI-PARTPHYS-289}
\title{125 GeV Higgs Boson, Enhanced Di-photon Rate, \\ and \\ Gauged $U(1)_{\rm PQ}$-Extended MSSM}

 \author{Haipeng An}
\affiliation{Perimeter Institute, Waterloo, Ontario N2L 2Y5, Canada}

\author{Tao Liu}
\affiliation{Department of Physics, University of California, Santa Barbara, CA 93106, USA}

\author{Lian-Tao Wang}
\affiliation{Enrico Fermi Institute, 
University of Chicago, Chicago, IL 60637, USA}
\affiliation{KICP and Dept. of Physics, Univ. of Chicago, 5640 S. Ellis Ave., Chicago IL 60637, USA}

\today

\begin{abstract}
The ATLAS and CMS collaborations have announced discovery of a $\sim125$ GeV Higgs boson, after a combined analysis of the di-photon and $ZZ$ search channels.  This observation has significant impact on low-energy supersymmetry. First, some fine-tuning is necessary to accommodate such a Higgs mass in the Minimal Supersymmetric Standard Model (MSSM) because the tree-level mass of the SM-like Higgs boson in the MSSM is relatively small. We study the possibility of lifting the mass of the SM-like Higgs boson by non-decoupling D-term from an additional $U(1)$ gauge symmetry. In particular, we focus on a gauged Peccei-Quinn symmetry which can also be related to a possible solution of the $\mu$ problem in the MSSM. In addition to the measurement of the mass of the Higgs, the data also reveals a tantalizing hint of a significantly enhanced di-photon signal rate, $1.56 \pm 0.43$ and $1.9\pm0.5$ times of the SM prediction in the CMS and ATLAS experiments, respectively. We demonstrate that such an enhancement can be accommodated in this MSSM extension. Anomaly cancellation requires the introduction of charged exotics. If some of them happen to be light and have sizable coupling to the SM-like Higgs boson, the di-photon signal rate can be enhanced significantly. EW precision measurements provide stringent constraints on this model. Taking these into account, we identify two benchmark scenarios. We argue that they are representative of large classes of viable models beyond our current example which can consistently enhance the Higgs to di-photon rate. We also comment on possible signals of such light exotics at the LHC.

\end{abstract}

\maketitle

\section{I Introduction}

Both the ATLAS and CMS collaborations have announced the discovery of a $\sim 125$ Higgs boson~\cite{CMS,ATLAS}, mainly based on  the combination of the di-photon and leptonic $ZZ$ Higgs searches at the $\sqrt{s}=7$ and 8 TeV LHC. 

This achievement fixes the last renormalizable parameter in the Standard Model (SM). At the same time, it carries important implications for new physics, particularly for low-energy supersymemmtry (SUSY). In the Minimal Supersymmetric Standard Model (MSSM), a SM-like Higgs boson with 125 GeV mass requires large corrections beyond the tree level prediction $(m_h)_{\rm tr} \approx m_Z \cos 2\beta$. Although such an 125 GeV SM-like Higgs boson can be accommodated, careful choices of parameters must be made~\cite{Carena:2011aa,Christensen:2012ei,Draper:2011aa,Cao:2012fz,Feng:2012jf,Hagiwara:2012mg,Christensen:2012si,Arbey:2012dq,Benbrik:2012rm}. Therefore, it is well motivated to consider possible extensions of the MSSM which give additional contribution to the mass of the SM-like Higgs boson at tree level~\cite{Benbrik:2012rm,Vasquez:2012hn,Ellwanger:2011aa,Hall:2011aa}. 

In this paper, we consider the possibility of enlarging the MSSM gauge symmetry by an additional $U(1)$. Such a gauge symmetry is quite generic in UV completions low energy supersymmetry, and many possible candidates have been proposed (for a review, e.g., see~\cite{Langacker:2008yv}). If its D-term is non-decoupling, it can provides non-trivial contribution to the tree-level mass of the SM-like Higgs boson~\cite{Batra:2003nj}.  In order to get a sizable correction, the Higgs field must be charged under this $U(1)$ symmetry. Moreover, the symmetry breaking scale of this extra U(1) symmetry cannot be much higher than the EW (EW) one.

We will focus on a gauged Peccei-Quinn (PQ) symmetry ($U(1)_{\rm PQ}$), under which by definition the Higgs fields carries non-trivial charges~\cite{Peccei:1977hh}. This can be connected to a possible solution of the $\mu$-problem~\cite{Kim:1983dt,Kim:2012at}, which is one of the central pieces of using low energy supersymmetry to address the hierarchy problem. 
The $\mu$ problem has its origin in a scale-violating term $\sim  \mu {\bf H_u H_d}$ in the superpotential of the MSSM, where ${\bf H_d}$ and ${\bf H_u}$ are down- and up-type Higgs supermultiplets. By introducing a spontaneously broken PQ global symmetry (a discrete version is the ${\bf Z_3}$ symmetry in the Next-to-MSSM), 
the bare $\mu$ term is forbidden and an effective one can be dynamically produced via
${\bf W_H} \sim \lambda {\bf S H_u H_d}$,
with $\mu_{\rm eff} = \lambda f_S$. Here ${\bf S}$ is a SM-singlet supermultiplet and $f_S$ is the vacuum expectation value (VEV) of its scalar component. 
Due to quantum gravity effects, it is expected that low energy symmetries should have their roots as gauge symmetries. 
In our case, we will further assume that such a gauge symmetry survives down to the TeV scale. In this paper, we will show that such a promotion for the PQ symmetry can significantly impact the Higgs physics. In particular,  the $U(1)_{\rm PQ}$ gauge symmetry introduces new D-terms which can raise the mass of the SM-like Higgs boson at tree level, enough to have $m_h = 125 $ GeV without significant radiative corrections. 

In addition to the mass of the Higgs boson, another interesting observation is that there is a possible excess in the di-photon signal with rate higher than the SM prediction. With a combined analysis of the $\sqrt{s}=7$ and 8 TeV LHC data, the CMS and the ATLAS collaborations obtain the best-fit signal strength: $1.56 \pm 0.43$ and $1.9\pm0.5$ times of the SM prediction, respectively\cite{CMS,ATLAS}. If such an excess is confirmed in the future, it would be an unambiguous indication for new physics.

There are two usual strategies to enhance the Higgs di-photon signal rate. The first one is by suppressing the width of its $b\bar b$ decay mode~\cite{Carena:1999xa,Ellwanger:2011aa}. In supersymmetric theories with two Higgs doublets, we have the coupling ratio $\frac{y_{hb\bar b}}{y_{hb\bar b}^{\rm SM}}= -\frac{\sin \alpha}{\cos \beta}$ for the SM-like Higgs boson (here we focus on the scenario with the lightest CP-even Higgs boson being SM-like. The discussion can be generalized to the case with the heavy CP-even Higgs boson being SM-like easily). Here the mixing angle $\alpha$ is defined as
\begin{equation}
\left( \begin{array}{c} {\rm Re} (H_u) \\ {\rm Re} (H_d) \\ \end{array} \right) = \frac{1}{\sqrt{2}} \left( \begin{array}{c} v_u + h \cos\alpha + H \sin\alpha \\ v_d - h \sin\alpha + H \cos\alpha \\ \end{array} \right) \ .
\end{equation}
$h$ and $H$ are light and heavy CP-even Higgs bosons, respectively. Suppressing the $h\to b\bar b$ decay width requires 
a small mixing angle for the SM-like Higgs boson, $i.e.$, $|\sin \alpha| < \cos \beta$. The SM-like Higgs boson therefore needs to be extremely up-type like for $\tan\beta > 1$. In the MSSM, it can only be achieved only through a cancellation between the tree-level and loop-level contributions to the off-diagonal Higgs mass term~\cite{Carena:2011fc}. But, in the extensions of MSSM, the quadratic terms of the Higgs sector can receive non-trivial corrections at tree level from new F-term or D-term corrections, so the mixing angle can be suppressed more easily~\cite{Azatov:2012wq,Ellwanger:2011aa}. A potential problem for this strategy is that the suppressed $b\bar b$ decay width can enhance the $ZZ$ and $WW$ signal rates as well. The current analyses by the CMS and ATLAS collaborations in these channels do not show such a feature, although one cannot completely rule out this possibility due to the limited statistics.

A more viable possibility is by enhancing the Higgs di-photon decay width.  Such a modification requires the existence of light charged exotics ($\sim 100 $ GeV), with large coupling to the SM-like Higgs boson.  The exotics carrying both color and EW charges might work. However, such particles typically bring larger contribution to the Higgs production via gluon fusion. More often than not, it would end up suppressing the rate of $pp \to h \to \gamma \gamma $~\cite{Carena:2011aa,Batell:2011pz,Cao:2012fz,Hagiwara:2012mg,Arhrib:2011vc,Arhrib:2012ia,Blum:2012kn}. In addition, direct searches at colliders have already put stringent lower bound on the mass of the colored exotics. Therefore, we will concentrate on the exotics with EW gauge charges only. There are several possibilities for the spin and coupling of the exotics. Probably the simplest case is the so-called Higgs portal couplings of the form $H^\dagger H Q^\dagger Q $, where $Q$ is some exotic scalar carrying electric charge.  It is well-known that (for recent discussions,  see, e.g., ~\cite{Batell:2011pz, Carena:2012xa}), in order to enhance the $h \gamma \gamma $ coupling significantly  and keep the coupling perturbative, the sign of this interaction must be negative. In supersymmetry, however, an interaction of this form is from the $F F^*$ term whose sign is positive. At the level of renormalizable couplings, therefore, the enhancement of the $h \to \gamma \gamma $ coupling in SUSY must come from Yukawa (trilinear) couplings between the Higgs boson and the exotic fermions (scalars). We also emphasize that the quantum number of such light exotics and the form of the couplings will be strongly constrained by the EW precision tests (EWPT). Remarkably, it is possible to satisfy these conditions in a special scenario of the MSSM with very light stau lepton and a large $h \tilde{\tau}_L \tilde{\tau}_R $ coupling~\cite{Carena:2011aa,Carena:2012gp}.

Interestingly, the gauged $U(1)_{\rm PQ}$ scenario contains the exotics which can enhance the $h \gamma \gamma$ coupling.  The $U(1)_{\rm PQ}$ symmetry is anomalous. Gauging it necessarily requires charged exotics to cancel its anomaly. The symmetries of the theory allow Yukawa and trilinear couplings between the SM-like Higgs boson and the exotics. If some of the exotics happen to be light, they can significantly enhance the Higgs di-photon decay width. Since the exotics usually can obtain their masses through the coupling with the $U(1)_{\rm PQ}$ breaking spurions, setting their masses to be light simply amounts to a choice of some dimensionless couplings. Precision EW tests can strongly constrain the possible parameters and the form of the couplings. Taking them into account, we identify two representative benchmark scenarios to illustrate the relevant Higgs physics. 

The rest of the paper is organized as follows. In Section II, we present the effective theory of the MSSM extension with a gauged $U(1)_{\rm PQ}$ symmetry. We discuss the mass of the SM-like Higgs boson in Section III. In Section IV, we give a general discussion of new physics contributions to the  Higgs to di-photon decay partial width, and its connection to EWPT observables. In Section V, we present an anomaly-free model and identify two benchmark scenarios where the experimental data can be fit correctly. Section VI contains our concluding remarks. In particular, we comment on the LHC signal  of the light exotics. We also argue that the two benchmark models identified in Section V are representative of large classes of viable models beyond our current framework which can consistently enhance the Higgs to di-photon rate.	  

\section{II Gauged $U(1)_{\rm PQ}$ Symmetry}

 A full model for the scenario considered in this paper needs to start from a sector which spontaneously breaks the $U(1)_{\rm PQ}$ symmetry. We assume that the PQ symmetry is spontaneously broken by the scalar components of the superfields ${\bf S_i}$, with their VEVs  being $\langle S_i \rangle = f_i$. The scalar potential of such a PQ breaking sector can be quite complicated. In this paper we will focus on a simple but instructive limit in which the $U(1)_{\rm PQ}$ symmetry breaking scale is somewhat larger than the scales of  EW symmetry breaking  and the soft SUSY breaking parameters,  $f_i > \Lambda_{\rm EW} \sim \Lambda_{
 \rm soft}$. In this case, we can integrate out the particles which become heavy after the $U(1)_{\rm PQ}$ symmetry breaking, in particular the ``radial modes" of the symmetry breaking fields. 

Since the SUSY breaking effect is smaller than the PQ symmetry one, it is convenient to group the light degrees of freedom in  an axion superfield 
\begin{eqnarray}
{\bf A}= A +\sqrt{2}\theta\tilde a +\theta^2 F_A\, , \ \ A = \frac{1}{\sqrt{2}}(s+i a) \ ,
\end{eqnarray}
with
${\bf S_i}=f_i e^{q_i {\bf A}/f_{\rm PQ}}$ in the representation of non-linear sigma model.
Here $f_{\rm PQ}=\sqrt{\sum_i q_i^2 f_i^2}$ is the $U(1)_{\rm PQ}$ breaking scale and $q_i$ is the $U(1)_{\rm PQ}$ charge of ${\bf S_i}$. 
At this stage, the axion ($a$) mass is protected by the Goldstone theorem, and it is related to the masses of the saxion ($s$), axino ($\tilde{a}$)  by SUSY. 
If the $U(1)_{\rm PQ}$ symmetry is global and SUSY breaking effect is relatively small, 
both $s$ and $\tilde a$ can be  light, with $m_s, m_{\tilde a} \sim m_a$. In this case,  the global $U(1)_{\rm PQ}$ theory provides a supersymmetric benchmark scenario of $\sim \mathcal O(1)$ GeV dark matter, with $\tilde a$ serving as the candidate~\cite{Draper:2010ew}. 
For a gauged $U(1)_{\rm PQ}$ symmetry, $a$ is eaten by the  $U(1)_{\rm PQ}$ gauge boson. 

The effective theory of the axion superfield is
\begin{eqnarray}
{\bf W_H}& =& \lambda {\bf S H_u H_d}  = \lambda f_S e^{ q_S {\bf A}/f_{\rm PQ}}  {\bf H_u H_d} \ ,  \nonumber \\
{\bf K} &=& \sum_i f_i^2 \exp\left(\frac{q_i{\bf(A+A^\dagger)}}{f_{\rm PQ}} + 2 g_{\rm PQ} q_i {\bf V_{\rm PQ}} \right) \nonumber \\ && + \sum_a {\bf H}_a^\dagger \exp(2 g_{\rm PQ} q_a {\bf V_{\rm PQ}}  + {\bf U_{\rm SM} }){\bf H}_a,  \label{10}
\end{eqnarray}
with $a$ summing over $\{u,d\}$ and $\bf U_{\rm SM}$ representing the contributions of the SM gauge symmetries. In our setup, $\lambda$ is a small parameter by assumption since $\Lambda_{\rm EW} \sim \mu_{\rm eff} = \lambda f_S  < f_{\rm PQ}$. 
We also include the SUSY breaking soft terms 
$V_{\rm soft} = - A_\lambda \lambda  S H_u H_d + {\rm h.c.} + \sum_a m_{H_a}^2 |H_a|^2 +  \sum_i m_{S_i}^2 |S_i|^2 $,  with their scales below  $f_{\rm PQ}$. 

Integrating out the saxion, we obtain a tree-level effective potential for the neutral Higgs sector 
\begin{eqnarray}\label{Higgs_potential}
V_{\rm WZ} &=& (|\mu_{\rm eff}|^2+m_{H_u}^2)|H_u|^2 + (|\mu_{\rm eff}|^2+m_{H_d}^2)|H_d|^2 \nonumber \\&&  - 2B_\mu {\rm Re} (H_u H_d) +\frac{1}{8}(g_2^2+g_Y^2)(|H_u|^2 - |H_d|^2)^2\nonumber \\
           && - g_{\rm PQ} q_{H_u}  \langle D_{\rm PQ} \rangle  (|H_u|^2 + |H_d|^2)  \nonumber  \\ && + a_1 |H_u H_d|^2  + a_2 (|H_u|^2 + |H_d|^2)^2  \nonumber  \\ && +  a_3 {\rm Re}(H_u H_d)(|H_u|^2 + |H_d|^2)  \ ,
\end{eqnarray}   
where the first two lines give the MSSM contributions, with $B_\mu = A_\lambda \mu_{\rm eff} $, and the other ones denote the leading corrections from the $U(1)_{\rm PQ}$ symmetry. 
In general, there are more corrections to the Higgs potential apart from listed in the last two lines of the above equation~\cite{Marandella:2005wc,Dine:2007xi}, and the coefficients of these terms are gauge dependent. A detailed discussion of the gauge choices is presented in Appendix A. Here we adhere to the Wess-Zumino gauge.        
$q_{\bf H_u} =  q_{\bf H_d}=-\frac{1}{2}q_{\bf S}$ has been assumed. To the order of $\lambda^2$, $a_1$, $a_2$ and $a_3$ are given by  
\begin{eqnarray}           
           a_1 &=& \left(\frac{2 q_{H_u} f_S}{f_{\rm PQ}}\right)^2\lambda^2 
                       \ \ \nonumber  \\
 a_2 &=&  g_{\rm PQ}^2 q_{H_u}^2 \left( \frac{1}{2} - \frac{ g_{\rm PQ}^2 f_{\rm PQ}^2 }{m_s^2} + \frac{4\lambda^2 f_S^2  }{ m_s^2}  \right)  \ ,  \nonumber   \\
  a_3 &=&\frac{-4 A_\lambda \lambda g_{\rm PQ}^2q_{H_u}^2  f_S }{ m_s^2}  \ .
 \label{coefficients}
\end{eqnarray}

$\langle D_{\rm PQ}\rangle$ is the VEV of the $U(1)_{\rm PQ}$ D-term, it is of the order soft SUSY breaking $\sim \Lambda_{\rm soft}^2$. The D-term contribution of the 3rd line of Eq.~(\ref{coefficients}) does not change the prediction of the Higgs  mass, since it just shifts the Higgs soft mass parameters $m_{H_{u,d}}^2$. In the  SUSY limit, we have $m_s^2 = m_{\tilde a}^2 = m_a^2 =m_{V_{\rm PQ}}^2= 2g_{\rm PQ}^2  f_{\rm PQ}^2$, and 
\begin{eqnarray}
         a_1 = \left(\frac{q_S f_S}{f_{\rm PQ}}\right)^2\lambda^2 ,    \ \ 
 a_2 =   \frac{2 q_{H_u}^2  \lambda^2 f_S^2}{ f_{\rm PQ}^2} , \ a_3=0  \ .
 \end{eqnarray}
It demonstrates the well-known result that the D-term contribution to the Higgs potential vanishes in the SUSY limit. 
The non-decoupling D-term contribution can be important if  the soft SUSY-breaking parameters of the PQ sector is not too small (but still below $f_{\rm PQ}$ by our assumption). In this case,  we have 
\begin{eqnarray}
a_1 &=&  {\mathcal O}(\lambda^2) \ , \ \  a_2  \ = \frac{1}{2}\ g_{\rm PQ}^2  q_{H_u}^2 \delta^2 +{\mathcal O}(\lambda^2) \ ,  \nonumber \\   
a_3 &=&\frac{-4 A_\lambda \lambda  g_{\rm PQ}^2 q_{H_u}^2  f_S }{ m_s^2}  + {\mathcal O}(\lambda^3) \ . 
\label{eq:a_nonsusy}
\end{eqnarray}
Here $m_s^2 =  2g_{\rm PQ}^2  f_{\rm PQ}^2 (1+\delta^2)$ with $\delta^2 = \frac{\sum_i m_{S_i}^2 q_i^2 f_i^2}{g_{\rm PQ}^2f_{\rm PQ}^4}$ representing the shift in $m_s^2$ induced by the softly SUSY breaking parameters $m_{S_i}^2$. In this case, the $U(1)_{\rm PQ}$ corrections to the MSSM Higgs via its D-term 
are dominant over the other sources. It provides a nice context to study the Higgs physics induced by a gauged $U(1)_{\rm PQ}$ symmetry.

This effective theory can also be built in super-unitary gauge, where the full  axion superfield is eaten by the $U(1)_{\rm PQ}$ vector superfield. 
In Appendix A, we present the effective Lagrangians in these two gauges. Although they have different forms, they are physically equivalent, leading to the same Higgs scattering amplitudes, vacuum energy and particle mass spectrum.

\begin{figure}[ht]
\begin{center}
\includegraphics[width=0.23\textwidth]{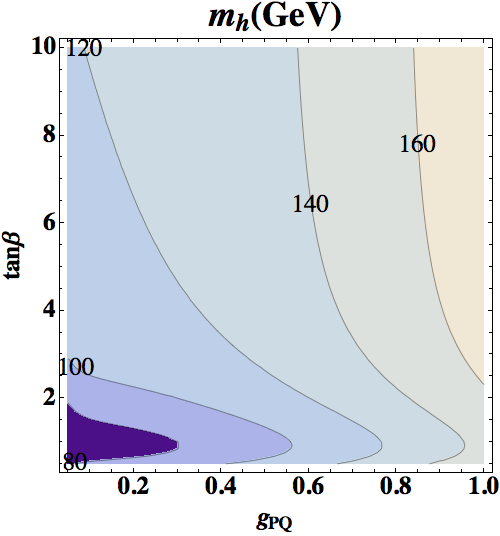}
\includegraphics[width=0.23\textwidth]{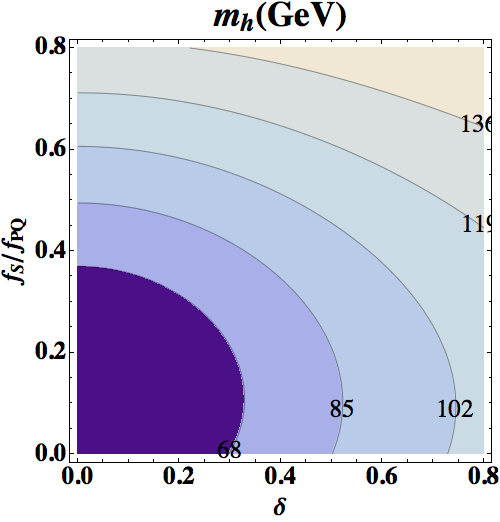}
\caption{$m_h$ contours with non-decoupling D-term contribution. For both plots, we assume \{$\lambda=0.3$, $\frac{A_\lambda}{f_{\rm PQ}}=0.1$\}. In addition, we set \{$\delta=0.6$, $\frac{f_S}{f_{\rm PQ}} =0.4$\} and \{$g_{\rm PQ}=0.6$, $\tan\beta=1$\} for the left and right panels, respectively. $\tan\beta=1$ provides the smallest tree-level Higgs mass, where the MSSM contribution is minimized. The loop corrections from stop and sbottom quarks have also been included, with the choice of the softly SUSY-breaking parameters: $A_t=A_b=\sqrt{m_{\tilde b}^2}=1200$ GeV, $\sqrt{m_{\tilde \tilde Q_3}^2}=\sqrt{m_{\tilde t}^2}=500$ GeV. }
\label{tree_mh}
\end{center}
\end{figure}

 \section{III Higgs Mass} 

The mass matrix for the CP-even Higgs bosons can be separated as
${\mathcal M}_H^2 = {\mathcal M}_{\rm MSSM}^2 + {\mathcal M}_{\rm PQ}^2$,
where  ${\mathcal M}_{\rm MSSM}^2$ denotes the MSSM contributions,  and ${\mathcal M}_{\rm PQ}^2$ denotes the $U(1)_{\rm PQ}$ corrections. We have  
\begin{eqnarray}
 ({\cal M}_{\rm PQ}^2/ v_{\rm EW}^2)_{11} &=& 2 a_2 \cos^2\beta + \frac{a_3}{4} (\frac{3}{2} \sin 2\beta - \sin^2 \beta \tan \beta)   \nonumber \\
 ({\cal M}_{\rm PQ}^2/ v_{\rm EW}^2)_{12} &=&  (\frac{a_1}{2}+a_2) \sin 2\beta + \frac{3}{4} a_3  \nonumber \\  
  ({\cal M}_{\rm PQ}^2/ v_{\rm EW}^2)_{21} &=& (\frac{a_1}{2}+a_2) \sin 2\beta + \frac{3}{4} a_3  \nonumber \\  
  ({\cal M}_{\rm PQ}^2/ v_{\rm EW}^2)_{22} &=&   2 a_2 \sin^2\beta + \frac{a_3}{4} (\frac{3}{2} \sin 2\beta - \cos^2 \beta \cot \beta)   \ ,  \nonumber
\end{eqnarray}            
with $\tan\beta = v_u /v _d$ and $\langle H_u^0 \rangle =  v_u / \sqrt{2}$, $\langle H_d^0 \rangle =  v_d / \sqrt{2}$, $v_{\rm EW} = (v_d^2 +v_u^2)^{1/2} = 246$ GeV. 
In the limit that the CP-odd Higgs boson is heavy, the lightest CP-even Higgs has a squared mass at tree level
\begin{eqnarray} \label{150}
(m_h^2)_{\rm tr} &\approx& m_Z^2 \cos^2 2 \beta   \\ && + \left (\frac{a_1}{2}  \sin^2 2\beta + 2 a_2 + a_3 \sin 2 \beta \right) v_{\rm EW}^2 \nonumber 
\end{eqnarray}
with the first term being the MSSM contribution. 

The variation of $m_h$ in the $g_{\rm PQ}-\tan\beta$ and $\delta-\frac{f_S}{f_{\rm PQ}}$ planes is shown in Fig.~\ref{tree_mh}, where the loop corrections from the MSSM  mediated by stop and sbottom quarks have been included. We see that $m_h=125$ GeV can be easily accommodated without heavy or split stops. The behavior of $m_h$ is mainly controlled by tree-level effects. With fixed $g_{\rm PQ}$, $m_h$ has a minimal value for $\tan\beta \sim 1$ where the MSSM tree-level contribution is minimized. If $\tan\beta$ is fixed, $m_h$ becomes larger as $g_{\rm PQ}$ increases. $\delta$ and $f_S/f_{\rm PQ}$ provide a set of measures of the D-term and F-term corrections to $m_h$. These features can be easily understood using Eq.~(\ref{150}) and Eq.~(\ref{coefficients}).

\section{IV $h \to \gamma\gamma$ And EW Precision Tests}

The effective Lagrangian of $h\rightarrow\gamma\gamma$ can be written as
\begin{eqnarray}
\label{eq:hgaga}
{\cal L}_{\rm eff} = -\frac{\alpha_{\rm em}}{2\pi} \frac{I}{v_{\rm EW}} F_{\mu\nu} F^{\mu\nu} h \ , 
\end{eqnarray}
where $I$ is a constant parameterizing the effective $h \gamma \gamma $ coupling.
Any particles coupled with it must get a mass from the Higgs VEVs. The effective coupling shown in Eq.~\ref{eq:hgaga} is induced by charged particles which can couple with the Higgs boson.  If the Higgs mass is smaller than that of the charged particles mediating the $h\gamma \gamma$ loop, the effective $h\gamma\gamma$ coupling can be calculated through the photon self-energy corrections~\cite{Ellis:1975ap,Shifman:1979eb}.  In the SM, there is only one Higgs doublet and the neutral component can be written as $H_{\rm SM} = (h+ v_{\rm EW})/\sqrt{2}$.  We have 
\begin{eqnarray}
I &=&   \sum_k\frac{b_{k}}{4}\frac{\partial}{\partial \log v_{\rm EW}} \log\left(\det{\cal M}_{k}^2\right) \label{333}
\end{eqnarray}
where for bosonic degrees of freedom ${\cal M}_k^2$ is the mass matrix and for fermionic degrees of freedom ${\cal M}_k^2 = {\cal M}_k^\dagger {\cal M}_k$, with $k$ running over all mass matrices of  charged bosonic and fermionic particles. $b_k$ is the coefficient in the beta function of QED, with  $b_{1/2} = \frac{4}{3} Q_f^2$ for Dirac fermion, $b_1 = -7 Q_v^2$ for charged vector boson, and $b_0 = \frac{1}{3} Q_S^2$ for charged scalar. In the SM, the main contributions come from top quark and $W$ boson. They have opposite signs. The $W$ boson contribution dominates over that of the top quark, and hence controls the sign of the $h\gamma\gamma$ coupling in the SM~\cite{Shifman:1979eb}. 

In the case of SUSY, there are two Higgs doublets, which makes the discussion complicated. The effective coupling can be modified to
\begin{eqnarray}\label{higgsI}
{I} &=& \sum_k\frac{b_{k}}{4}\left[\cos\alpha\frac{\partial}{\partial v_u} \log\left(\det{\cal M}_{k}^2\right)\right.\nonumber \\
&&- \left.\sin\alpha\frac{\partial}{\partial v_d} \log\left(\det{\cal M}_{k}^2\right)\right]  \ .
\end{eqnarray}
Usually there is no fixed relation between the Higgs alignment $\alpha$ and VEV alignment $\beta$. In this paper, we concentrate on the scenarios in which the decays of the SM-like Higgs boson to $WW$,  $ZZ$, $b\bar b$,  $ \tau \bar \tau$ channels are similar to that of the SM Higgs, and only the loop-dominant $h\rightarrow\gamma\gamma$ channel is modified. In comparison with the SM, the decay amplitude of $h\rightarrow b\bar b$ and $h\rightarrow WW/ZZ$ in MSSM are scaled by a factor of $\frac{\cos\alpha}{\sin\beta}$ and $\sin(\beta-\alpha)$, respectively. If $\beta-\alpha\approx\pi/2$, their decay widths are approximately equal to the SM ones. The contributions from the $W$ boson and top quark to $h\rightarrow\gamma\gamma$ is also similar to the SM one, which give $I_W\approx-2.1$ and $I_{\rm top}\approx0.5$. The effective Lagarangian is reduced to the one in Eq. (\ref{333}).

If the masses of the intermediate particles are smaller than half of the Higgs mass, they can be pair-produced via the Higgs decay. Treating the exotic induced $h\to \gamma\gamma$  coupling as point-like, as described by the effective Lagrangian in Eq.~(\ref{eq:hgaga}), is not appropriate.   For $m_h=125$ GeV, the current lower bounds on the mass of the charged new particles are around 100 GeV~\cite{Achard:2003ge}. Therefore, the effective theory description is always valid. For relatively light mediators, there are corrections to $b_k$'s up to the order of $m_h^2/4 m_{\rm mediator}^2$. These corrections are small and will not change the conclusions reached in this section qualitatively. The exact formulae can be found in Ref.~\cite{Shifman:1979eb,CP}, and they will be used in our numerical calculations in the next section.

From Eq.~(\ref{higgsI}), we can see that the sign of the effective coupling is determined by two factors, the $\beta$-function $b_k$ and the derivatives of the mass matrix with respect to the Higgs VEVs. For fermions and complex scalars, $b_k$ is positive. In the SM, the fermions have ${\cal M}^\dagger {\cal M} \sim |y|^2 v_{u,d}^2$, where $y$ is Yukawa coupling. Therefore, the derivative is always positive. For general exotic fermions, however, there can have vector-like mass terms.  The derivative can be either positive or negative. For exotic scalars, in SUSY models, generally they can obtain mass from the Higgs VEVs in two ways. One is from the F-term of the exotic superfield. In this case the derivatives are always positive. The other is from the $A$-term between the exotics and the Higgs boson. In this case, the derivatives can be either positive or negative. In the next section, we will discuss benchmark models of these two scenarios. 

To enhance the $h\to \gamma\gamma$ signal rate significantly, the contribution from the exotic states needs to be comparable with the SM contributions from $W$ boson and top quark. As a result, either the fermion or the scalar mediators should be light. Since these particles carry EW charges, they may non-trivially contribute to the observables of the EWPT. Therefore, the EWPT can  provide strong constraints on the models discussed here. These potential contributions include oblique and non-oblique corrections. We will focus on the former, because the non-oblique ones are sensitive to the couplings of the mediators with the SM fermions, which can be taken to be small.  The oblique corrections represent new physics effects in the vacuum polarization of the SM gauge bosons, and are usually parameterized by the Peskin-Takeuchi parameters, $S$, $T$ and $U$~\cite{Peskin:1990zt}. The $U$ parameter is not very sensitive to new physics, and only receives contributions from dimension-eight operators or above, so we will not discuss it in the following analysis. The $U(1)_{\rm PQ}$ gauge boson can also have non-trivial contributions to the EWPT observables, via its mixing with the $Z$ boson. This has been discussed extensively (e.g., see~\cite{Erler:2009jh}). For the scenarios discussed in this paper, the $Z-Z'$ mixing is small because the $U(1)_{\rm PQ}$ scale is relatively high while its gauge coupling is not very large. Such contributions therefore are well under control.

\begin{table}[t]
\begin{center}
\begin{tabular}{|c|c|c|c|}
\hline
 Particles & Gauge charges  & Particles &  Gauge charges \\
 \hline
\hline 
${\bf L_i}$ & (1; 2; $-1/2$; $1/2$) &
${\bf Q_i}$ & (3; 2; 1/6; $1/2$) \\
\hline
${\bf \bar N_i}$ & (1; 1; 0; $1/2$) & ${\bf \bar u_i}$ & ($\bar 3$; 1; $-2/3$;  $1/2$)\\
\hline
${\bf \bar e_i}$ & (1; 1; 1; $1/2$) & ${\bf \bar d_i}$ & ($\bar 3$; 1; 1/3;  $1/2$)\\
\hline 
${\bf H_d}$ & (1; 2; $-1/2$; $-1$) & ${\bf H_u}$ & (1; 2; $1/2$; $-1$) \\
\hline
\hline
${\bf T_1}$ &  (3; 1; $2/3$; $-1$) & $ {\bf T_1^c}$ &  ($\bar 3$; 1; $-2/3$; $-1$)  \\
\hline
${\bf T_2}$ &  (3; 1; $2/3$; $-1$) & $ {\bf T_2^c}$ &  ($\bar 3$; 1; $-2/3$; $-1$)  \\
\hline
${\bf T_3}$ &  (3; 1; $-1/3$; $-1$) & $  {\bf T_3^c}$ &  ($\bar 3$; 1; $1/3$; $-1$)  \\
\hline
${\bf D_1}$ &  (1; 2; $1/2$; $-1$) & $  {\bf D_1^c}$ &  (1; 2; $-1/2$; $-1$)  \\
\hline
${\bf D_2}$ &  (1; 2; $1/2$; $-1$) & $  {\bf D_2^c}$ &  (1; 2; $-1/2$; $-1$)  \\
\hline
$ {\bf  X}$ & (1; 1; 1; $2$) & $ {\bf X^c}$ & (1; 1; -1; $2$) \\
\hline
$ {\bf  N}$ & (1; 1; 0; $2$) & $ {\bf N^c}$ & (1; 1; 0; $2$) \\
\hline
\hline
${\bf S }$ & (1; 1; 0; $2$) &${\bf S^c}$ & (1; 1; 0; $-2$)  \\
\hline
${\bf S_1}$ & (1; 1; 0; $-4$) &${\bf S_1^c}$ & (1; 1; 0; $4$)  \\
\hline
${\bf S_2}$ & (1; 1; 0; $-2$) & &   \\
\hline
\end{tabular}
\end{center}
\caption{Particle content in a supersymmetric model with anomaly-free $U(1)_{\rm PQ}$ gauge symmetry.} \vspace{5 mm}
\label{table1}
\end{table}

\begin{table}[t]
\begin{center}
\resizebox{85mm}{!}{
\begin{tabular}{|c|c|c|c|c|}
\hline
 $g_{\rm PQ}$ & $f_{\rm PQ}$ (GeV) & $f_S/f_{\rm PQ}$ & $A_\lambda/f_{\rm PQ}$ &  $\lambda$ \\
 \hline
0.6 & 2500 & 0.4 & 0.1 & 0.3  \\    
\hline
 $\tan\beta$ & $\delta$ & $A_{\tilde \gamma}$  (GeV) & $A_{\bar \gamma}$  (GeV)& $\tilde \gamma, \bar \gamma$  \\        
\hline
1.3 & 0.6 & $0$ & 0  & 1.6   \\
\hline
$m_D$  (GeV) & $m_X$ (GeV) & $m_{\tilde D, \tilde X, \tilde N}^2$ (GeV$^2$)  & $A_{\tilde t}$ (GeV) & $m_{\tilde Q_3}^2, m_{\tilde t}^2$  (GeV$^2$)  \\        
\hline
440 & 330 &  $1000^2$  & 1200 &  $500^2$ \\
\hline
$a_1$ & $a_2$  &$a_3$ & $B_\mu$ (GeV$^2$) & $\mu_{\rm eff}$ (GeV)   \\        
\hline
0.06 & 0.09 & $-0.02$   & $7.5\times 10^4$ &  300 \\
\hline
 $m_h$ (GeV) & $m_{\psi_1^c}$ (GeV)  &$m_{\psi_1^0}$  (GeV) &  $m_{\phi_1^c}$ (GeV) & $m_{\phi_1^0}$ (GeV)   \\        
\hline
 125 &  105  & 105 & 943 & 943   \\
 \hline
  $R(h\to \gamma\gamma)$ & $\Delta S$ & $\Delta T$ & & \\
\hline
$1.8$ &0.11& 0.10&& \\
\hline
\end{tabular}
}
\end{center}
\caption{Benchmark scenario I, where the di-photon signal rate is mainly enhanced by light fermion spectators. $M_{D, X}$ are vector-like masses of the $({\bf D_p, D_p^c})$ and $({\bf X, X^c})$ fermion components, $m_{\tilde D, \tilde X, \tilde N}$ are the soft mass parameters of their scalar components (including $({\bf N, N^c})$), $X_t$ and $M_{\tilde t}$ are the mixing and soft mass parameters of stop quarks. $m_{\psi_1^c}$ and $m_{\psi_1^0}$ are the masses of the lightest charged and neutral exotic fermions, respectively. $m_{\phi_1^c}$ and $m_{\phi_1^0}$ are the masses of the lightest charged and neutral exotic scalars, respectively. We have also defined $R(h\to \gamma\gamma)= \sigma^{\gamma \gamma}_{U(1)_{\rm PQ} \rm MSSM} /  \sigma^{\gamma \gamma}_{\rm SM}$.}\vspace{7 mm}
\label{table2}
\end{table}

\begin{table}[t]
\begin{center}
\resizebox{85mm}{!}{
\begin{tabular}{|c|c|c|c|c|}
\hline
 $g_{\rm PQ}$ & $f_{\rm PQ}$ (GeV) & $f_S/f_{\rm PQ}$ & $A_\lambda/f_{\rm PQ}$ &  $\lambda$ \\
 \hline
0.6 & 2500 & 0.4 & 0.1 & 0.3  \\    
\hline
 $\tan\beta$ & $\delta$ & $A_{\tilde \gamma}$  (GeV) & $A_{\bar \gamma}$  (GeV)& $\tilde \gamma, \bar \gamma$   \\        
\hline
6 & 0.6 & $1440$ & 1000  & 0.5    \\
\hline
 $m_D$  (GeV) & $m_X$ (GeV) & $m_{\tilde D, \tilde X, \tilde N}^2$ (GeV$^2$) & $A_{\tilde t}$ (GeV) & $m_{\tilde Q_3}^2, m_{\tilde t}^2$  (GeV$^2$)  \\        
\hline
500 & 350 &  $100^2$  & 1200 &  $500^2$ \\
\hline
$a_1$ & $a_2$  &$a_3$ & $B_\mu$ (GeV$^2$) & $\mu_{\rm eff}$ (GeV)   \\        
\hline
0.06 & 0.07 & $-0.02$   & $7.5\times 10^4$ &  300 \\
\hline
 $m_h$ (GeV) & $m_{\psi_1^c}$ (GeV)  &$m_{\psi_1^0}$  (GeV) &  $m_{\phi_1^c}$ (GeV) & $m_{\phi_1^0}$ (GeV)   \\        
\hline
 125 &  325  & 325 & 104 & 233   \\
 \hline
  $R(h\to \gamma\gamma)$ & $\Delta S$ & $\Delta T$ & & \\
\hline
$1.7$ &0.03& 0.08&& \\
\hline
\end{tabular}
}
\end{center}
\caption{Benchmark scenario II, where the di-photon signal rate is mainly enhanced by light scalar spectators. } \vspace{7 mm}
\label{table3}
\end{table}

\begin{figure}[ht]
\begin{center}
\includegraphics[width=0.21\textwidth]{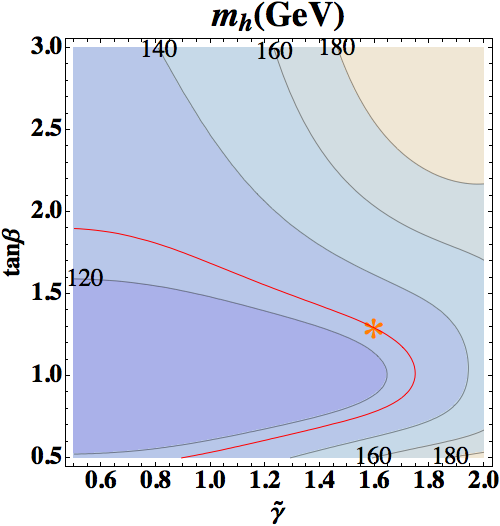}
\includegraphics[width=0.21\textwidth]{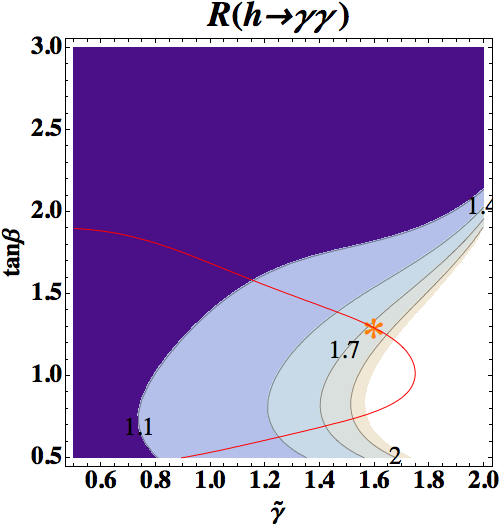}
\includegraphics[width=0.21\textwidth]{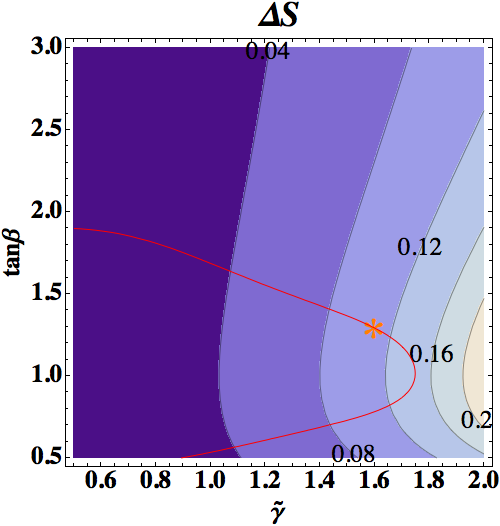}
\includegraphics[width=0.21\textwidth]{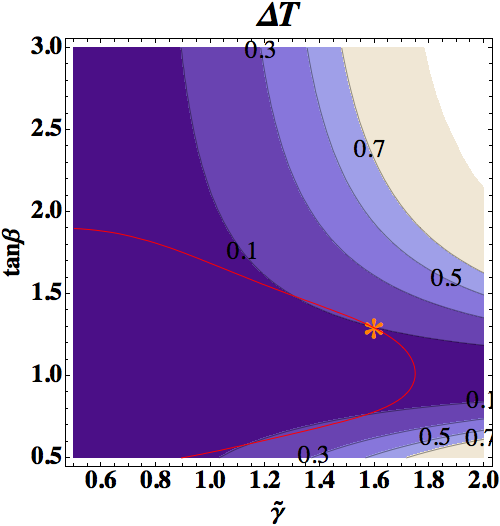}
\caption{$m_h$, $R (h\to \gamma\gamma)$, $\Delta S$ and $\Delta T$ contours in benchmark scenario I. Red line gives the $m_h=125$ GeV contour. The yellow stars correspond to benchmark scenario I, assuming for simplicity $\gamma_{X}^p=\gamma_{N}^p=\gamma_{X^c}^q=\gamma_{N^c}^q= \tilde \gamma$, parameters other than $\tilde\gamma$ and $\tan\beta$ are fixed as shown in Table~\ref{table2}. The red curves show the region where $m_h = 125$ GeV. }
\label{contour1}
\end{center}
\end{figure}

\begin{figure}[ht]
\begin{center}
\includegraphics[width=0.21\textwidth]{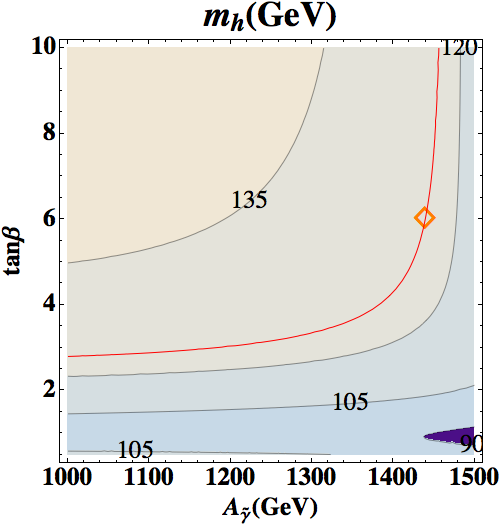}
\includegraphics[width=0.21\textwidth]{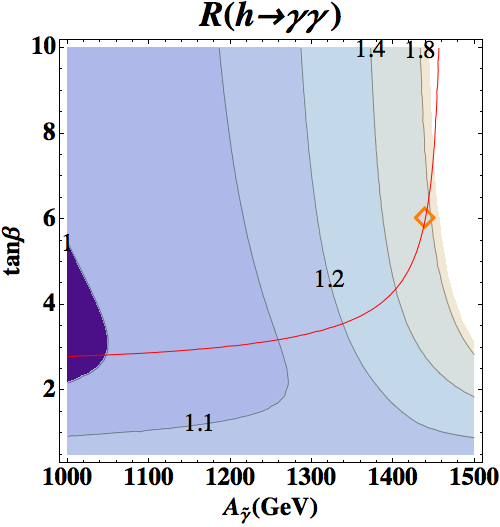}
\includegraphics[width=0.21\textwidth]{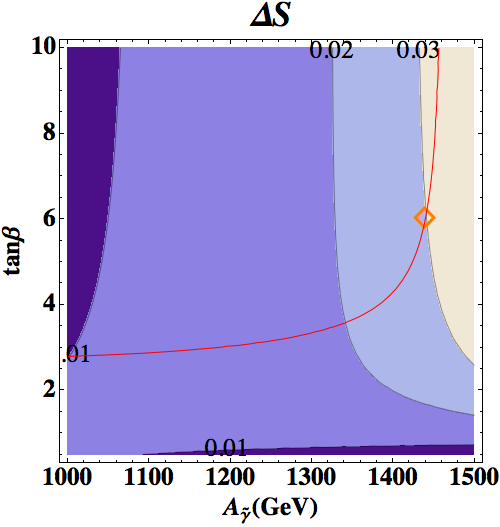}
\includegraphics[width=0.21\textwidth]{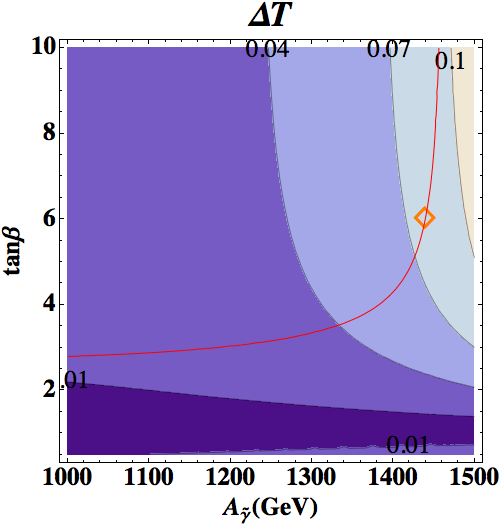}
\caption{$m_h$, $R (h\to \gamma\gamma)$, $\Delta S$ and $\Delta T$ contours in benchmark scenario II. Red line gives the $m_h=125$ GeV contour, parameters other than $A_{\tilde\gamma}$ and $\tan\beta$ are fixed as shown in Table~\ref{table3}. The yellow stars correspond to benchmark scenario II. The red curves show the region where $m_h = 125$ GeV.}
\label{contour2}
\end{center}
\end{figure}

\section{V An Example}

In this section, we construct an explicit model which realizes the ideas discussed in the previous sections. $U(1)_{\rm PQ}$ with only the SM matter content is anomalous. 
There are many possible choices of spectators to cancel this anomaly. However, new EW doublets and color triplets carrying $U(1)_{\rm PQ}$ charges are always required for canceling the $SU(2)^2 \times U(1)_{\rm PQ}$ and $SU(3)^2 \times U(1)_{\rm PQ}$ anomalies, respectively. 
As an illustration, let us consider a model given in Table~\ref{table1}. Other possible implementations have qualitatively similar features. The superpotential of this model is given by  
\begin{eqnarray}
{\bf W} & =&  {\bf W_{\rm H}}  \nonumber \\ && +  \delta_N {\bf S_1 N N^c}+ \beta^{pq} {\bf S D_p D_q^c} +  \delta_X {\bf S_1 X X^c}  \nonumber \\
 && +  \gamma_{X^c}^p {\bf H_u D_p X^c} +\gamma_{N^c}^p {\bf H_d D_p N^c}  \nonumber \\ && + \gamma_{X}^q {\bf H_d D_q^c X}+ \gamma_{N}^q {\bf H_u D_q^c N }  \nonumber \\
&& +{\bf W_{\rm Y}} ({\bf H_u \leftrightarrow D_k, H_d \leftrightarrow D_k^c })  \nonumber \\
&& +{\bf W_{\rm LQ}} +{\bf W_{\rm S}}
\label{eq:Wmodel}
\end{eqnarray}
with $p, q=1, 2$. ${\bf W_{\rm H}}$ is presented in Eq. (\ref{10}).   We have only displayed explicitly the interactions most relevant for our discussion on the modification of the $h \gamma \gamma$ effective coupling.  ${\bf W_{\rm Y}}$ has the same form as the MSSM Yukawa couplings, but with the replacements ${\bf H_u \leftrightarrow D_p}$ and  ${\bf H_d \leftrightarrow D_q^c}$.  ${\bf W_{\rm S}}$ includes all terms with only the PQ symmetry breaking fields $({\bf S, S^c, S_1, S_1^c, S_2})$. 
${\bf W_{\rm LQ}}$ contains the interactions involving the color triplets ${\bf T_{1,2,3}^{(c)}}$. It has the coupling of the form $ {\bf S T T^c}$, and  $\kappa^{rs} {\bf L_r Q_s T_3 } + \kappa^{rsp} {\bf \bar N_r \bar u_s T_p}$ with $r, s=1, 2, 3$.
After the $U(1)_{\rm PQ}$ symmetry is broken, $({ \bf T_r, T_s^c})$ and  $({\bf D_p, D_q^c})$ obtain vector-like masses by coupling with the supermultiplet ${\bf S}$.
An important feature is that $({ \bf T_r, T_s^c})$ and $({\bf D_p, D_q^c})$ decay into the SM particles via the interactions in ${\bf W_{\rm LQ}}$ and ${\bf W_{\rm Y}}$, respectively. This can help avoid the overproduction of the exotic particles in the Universe.
From the choice of the $U(1)_{\rm PQ}$ charges displayed in Table~\ref{table1}, the only renormalizable couplings between the color triplets and the Higgs bosons are of the form ${H \tilde{q}  \tilde{\ell} \tilde{T}}$ or ${H H  \tilde{T} \tilde{T}}$. ${H \tilde{q}  \tilde{\ell} \tilde{T}}$ will not contribute to the Higgs decay directly. Furthermore, they can be suppressed by choosing small leptoquark-type couplings. ${H H  \tilde{T} \tilde{T}}$ will enter $h gg$ effective coupling. However, this contribution can be small if the color triplets is heavy, $m_{\rm T} \sim $ TeV.  For these reasons, we will assume that the triplets are heavy and would not discuss them further in this section.

Since the EWPT may give strong constraints on our scenario, we begin with a more general discussion on this issue before presenting the benchmark models. 
The T parameter is a measure of the breaking of the custodial symmetry, $SU(2)_c$. Apart from the mixing between $Z$ and $Z'$, in general, there are two contributions which break $SU(2)_c$ explicitly. One is from the Yukawa coupling between the exotics and the Higgs fields in the superpotential, such as the difference between the ${ \bf H_u D_p X^c}$ and ${\bf H_d D_p N^c}$ couplings, and the corresponding A-terms. In this paper, we  choose to preserve the explicit $SU(2)_c$ by choosing the relevant Yukawa couplings to be equal. Therefore, instead of four independent Yukawa couplings, we have two, $\gamma_{X^c}^p=\gamma_{N^c}^p=\tilde \gamma$ and $\gamma_{X}^q=\gamma_{N}^q=\bar \gamma$, and for the corresponding A-terms, we have $A_{X^c}^p=A_{N^c}^p=A_{\tilde \gamma}$ and $A_{X}^q=A_{N}^q=A_{\bar \gamma}$. We note that certain fine-tuning is necessary in making this choice, as these couplings also receive one-loop corrections proportional to the explicit $SU(2)_c$ breaking SM Yukawa couplings. 
Another contribution to the T parameter comes from  the difference between $v_u$ and $v_d$, which breaks $SU(2)_c$ spontaneously. Of course, we can find  parameter space where the two contributions cancel with each other. To avoid severe fine-tuning, the bulk of parameter space with such cancellation should not have large explicit or spontaneous $SU(2)_c$ violation. Therefore, it should not be very different from the limit that we are considering. 

Two benchmark points are presented in Table~\ref{table2} and Table~\ref{table3}. The parameter $\delta$, defined right after Eq.~(\ref{eq:a_nonsusy}), controls the corrections to the Higgs sector via the $U(1)_{\rm PQ}$ D-terms and is assumed to be sizable for both benchmark points. In the first benchmark, the lightest charged ($\psi_1^c$) and neutral ($\psi_1^0$) fermion spectators are light. They have a large coupling to  the Higgs fields, for enhancing the di-photon signal rate. To get a small $\Delta T$, the spontaneous breaking needs to be small. Hence,  $\tan\beta\approx1$ is required. 
 For the second benchmark point, where one of the charged scalar is light and couples to the Higgs fields with large $A$ terms (again for enhancing the di-photon signal rate),
there is an accidental cancelation which leads to $\Delta T$ to be a few times smaller than its natural value.  A detailed discussion of this accidental cancelation can be found in Appendix B. Therefore, even in the case of large $\tan\beta$, the $T$ parameter can still be within the experimental limit. 

In addition, although the doublets $({\bf D_p, D_q^c})$ are vector-like under the EW gauge symmetry and their fermionic components have a degenerated mass spectrum, the $S$ parameter can still receive non-zero corrections. This is because $({\bf D_p, D_q^c})$ mix with $({\bf X, X^c})$ and  $({\bf N, N^c})$,  while the latter violate the weak isospin (recall that the $S$ parameter preserves the custodial but measures the  violation of  the weak isospin). 

The dependence of $m_h$, $R(h\to \gamma\gamma)$, $\Delta S$ and $\Delta T$ on $\tilde \gamma = \bar \gamma$ and $\tan\beta$ in the first benchmark is shown in Fig.~\ref{contour1}. 
As is expected, with a fixed $\tan\beta$, $R(h\to \gamma\gamma)$ tends to be enhanced for a larger $|\tilde \gamma|$; and with a fixed $\tilde \gamma$, $\Delta T$ tends to be smaller while $\tan\beta$ is close to 1. From the right plot on the first row of Fig.~\ref{contour1}, we can see that the correction to the rate of $h\rightarrow\gamma\gamma$ also has a maximum at $\tan\beta = 1$ with fixed $\tilde \gamma$. We can understand this from the form of the mass matrix of charged fermionic exotics, 
\begin{equation}\label{fermionmass}
{\cal M}_f \sim \left( \begin{array}{cc} M_D ~&~ \tilde \gamma v_u \\ \tilde \gamma v_d ~&~ M_X \\\end{array} \right) \ ,
\end{equation}
where $M_D$ and $M_X$ are the vector-like mass for the doublet and charged singlet exotics, respectively. Then, we can get 
\begin{equation}
\left( \cos\alpha\frac{\partial}{\partial v_u} - \sin\alpha\frac{\partial}{\partial v_d} \right) \det{\cal M}_f \sim - \tilde \gamma^2 v_{\rm EW} \cos(\alpha+\beta) \ .
\end{equation}
Since we are interested in the region where $\alpha \approx \beta - \pi/2$, we have $\cos(\alpha+\beta)\approx \cos (2\beta - \pi/2)$, which peaks at $\beta = \pi/4$. Therefore, from Eq.~(\ref{higgsI}), we can see that the correction to $I$ reaches its maximum at $\tan\beta = 1$. 

The dependence of $m_h$, $R(h\to \gamma\gamma)$, $\Delta S$ and $\Delta T$ on $A_{\tilde \gamma}$ and $\tan\beta$ in the second benchmark is shown in Fig.~\ref{contour2}. In this region, since the contribution from $A$ terms dominates over the ones from the Yukawa couplings, the loop correction to Higgs mass from exotics is negative. Therefore, from the first plot in Fig.~\ref{contour2}, we can see that $m_h$ becomes smaller with larger $A_{\tilde\gamma}$. For the correction to the rate of $h\rightarrow\gamma\gamma$, in the region where $A_{\tilde\gamma}^2 > A_{\bar\gamma}^2$, the charged scalar exotics mainly couple to the Higgs through $H_u$ and therefore the correction to the $h\rightarrow\gamma\gamma$ rate behaves similarly to up-type quarks. As a result, the relevant part of the mass matrix of the charged exotics in this limit can be reduced to 
\begin{equation}
{\cal M}_s^2 \sim \left( \begin{array}{cc} M_D^2 + m_D^2 ~&~  A_{\tilde\gamma} v_u \\ A_{\tilde\gamma}^\dagger v_ u ~&~ M_X^2 + m_X^2 \\ \end{array} \right) \ ,
\end{equation}
from which we can get 
\begin{equation}
\left( \cos\alpha\frac{\partial}{\partial v_u} - \sin\alpha\frac{\partial}{\partial v_d} \right) \det{\cal M}_f^2 \sim - |A_{\tilde\gamma}|^2 v_{\rm EW} \sin^2\beta \ ,
\end{equation}
where the relation $\cos\alpha \approx\sin\beta$ is used. Therefore, we can see that in this region the correction to the rate of $h\rightarrow\gamma\gamma$ goes up slowly with $\tan\beta$ which is shown in the region $A_{\tilde\gamma}$ around 1500 GeV in Fig.~\ref{contour2}, where $A_{\bar\gamma}$ is fixed to be 1000 GeV. In the region $A_{\tilde\gamma} = A_{\bar\gamma}$ the dependence of the correction to $h\rightarrow\gamma\gamma$ on $\tan\beta$ is more complicated, and numerical simulation shows that the dependence is not monotonic, as shown in Fig.~\ref{contour2}.  
For the contributions to the $T$ parameter, because of the accidental cancelation discussed in Appendix B, $\Delta T$ is typically small, while $\tan\beta \sim 1$ can bring a further suppression. 

\section{VI Concluding Remarks}

Motivated by the discovery of a $\sim 125$ GeV Higgs boson in the CMS and ATLAS experiments, we study the possibility of lifting  the tree-level mass of the SM-like Higgs boson in MSSM extended by an extra $U(1)$ gauge symmetry. For definiteness, we focus on the scenarios with a gauged $U(1)_{\rm PQ}$ symmetry which can also be connected to a possible solution to the $\mu$ problem in the MSSM. We limit ourselves to the parameter region in which the  softly SUSY-breaking scale is below the $U(1)_{\rm PQ}$ breaking scale, $f_{\rm PQ} > \Lambda_{\rm soft} \sim \Lambda_{\rm EW}$. In this case, we can work in the framework of effective theory with only the axion supermultiplet.  We explicitly demonstrate that non-trivial corrections of the $U(1)_{\rm PQ}$ D-term to the Higgs physics necessarily require sizable softly SUSY breaking effects in the PQ symmetry-breaking sector. In particular, a correction $\sim \mathcal O(10)$ GeV to the Higgs mass at the tree level can be achieved for $f_{\rm PQ} \sim \mathcal O(1)$ TeV and $\frac{\Lambda_{\rm soft}}{f_{\rm PQ}} \sim \mathcal O(0.1)$. In addition to the Higgs mass, the LHC data also reveals a tantalizing hint of a significantly enhanced di-photon signal rate.  We show that this feature can be accommodated in this scenario. Gauging the $U(1)_{\rm PQ}$ symmetry necessarily requires the charged exotics for anomaly cancellation. If they happen to be light and have sizable couplings with the Higgs boson, the di-photon signal rate can be significantly enhanced. With the bounds from the EWPT considered, we identify two benchmark scenarios where a light charged exotic fermion and scalar plays a crucial role in modifying the $h \gamma \gamma$ effective coupling, respectively.

Testing this scenario at the LHC is relatively difficult. The first signal is probably still from the modified decays of the SM-like Higgs boson. A confirmation of the enhanced  di-photon signal rate would provide a potential evidence  for this scenario. In this case, it is obviously important to search for the light charged exotic mediators directly at the LHC. These exotics can be produced through the processes of weak interaction and then decay into the SM particles via the interactions described in the superpotential $\bf W_Y$. Their signals are similar to that of the Higgsinos, or  sleptons.  With the accumulation of the data, we should be able to probe these exotics. Of course, it is also possible to search for the leptoquark-type exotics $\bf T_{1,2,3}$, which are required for cancelling the $SU(3)^2\times U(1)_{\rm PQ}$ anomaly.  Although not directly related to the modification of the Higgs di-photon decay channel, these colored particles should not be too heavy since their mass is generated through PQ symmetry breaking. They can be pair-produced via QCD processes at the LHC, with leptoquark-like signals. We will leave these studies to a future work.

Although our study is mainly done in the context of a specific model, the lessons we have learned are fairly general. The two benchmarks are representatives of large classes of models in which an enhancement of the $h\to \gamma\gamma$ signal rate does not lead to a violation of the EWPT constraints. The first benchmark contains four $SU(2)_L$ doublets and two charged singlets for the sake of anomaly cancelation. This is only a little larger than the minimal model.  As discussed in Sec.~IV, for the corrections from fermionic exotics to have the same sign as the W boson contribution, the mass of the fermionic exotics must have two sources. One is chiral, coming from Yukawa like couplings with Higgs fields.  At the same time, a Dirac mass term is necessary. Therefore, we at least need to introduce two doublets.
Moreover, to avoid explicit breaking of the custodial $SU(2)_c$, we must at least introduce two singlets.
The mass matrices of the fermionic charged exotics can always be written in the form of Eq.~(\ref{fermionmass}). Hence, the discussions for the correction to Higgs to di-photon decay rate made in Sec.~V are in general applicable. 
The second benchmark is a very special case in which a charged scalar can do the job. Another similar example in this class is the light stau scenario in the MSSM~\cite{Carena:2011aa}. We see that, due to the constraints from the EWPT, a certain amount of fine tuning is unavoidable in models with an enhanced $h \gamma \gamma$ effective coupling. This implies that if such an enhancement is confirmed, it would point us to a new direction of model building not completely guided by the reduction of such fine-tuning. If only the $h\to \gamma\gamma$ signal rate is enhanced while the other ones are not modified, the new exotic states will carry the EW quantum numbers only. Their collider signals are similar to that of the EW-ino and the slepton. Hence, it would be  challenging to search, unless they are part of a long decay chain starting with some colored states. 

\begin{center}
{\bf Acknowledgments}
\end{center}  

\vspace*{-5pt}

H.A. is supported by the Government of Canada through NSERC and by the Province of Ontario through MEDT.
T.L. is supported by the DOE Grant DE-FG02-91ER40618 at U.~California, Santa Barbara.   L.T.W. is supported by the NSF under grant PHY-0756966 and the DOE Early Career Award under grant DE-SC0003930. T.L. would thank D. Morrissey for useful discussions.

\appendix

\section{A Comparison between Wess-Zumino Gauge and Super-Unitary Gauge}
\label{app:gauge}

In this section, we calculate the corrections to the MSSM Higgs potential in two different choice of gauge, which are the Wess-Zumino gauge and the Super-Unitay gauge. We show that although in the two choices the corrections to the Higgs potential are different as well as the corrections to the VEVs of Higgs fields. But the physical observables are the same in the two gauge choices. 

\subsection{A.1 Effective Higgs Potential: Wess-Zumino Gauge}

In the Wess-Zumino gauge, the effective potential for the neutral Higgs sector can be obtained from ${\bf K}$, ${\bf W_{\rm H}}$ and $V_{\rm soft}$ which are defined in Eq.(\ref{10}) or its below.  
It is easy to get
\begin{eqnarray}
V_{\rm WZ} &\sim&  \frac{g_{\rm PQ}^2}{2} \left[ \sum_i q_i f_i^2 (X^\dagger X)^{q_i} + q_{H_u} |H_u|^2 + q_{H_d} |H_d|^2\right]^2 \nonumber \\
&& + \lambda^2 f^2_S (X^\dagger X)^{q_S} ( |H_u|^2 + |H_d|^2) \nonumber \\ &&+ \lambda^2  \frac{q_S^2 f_S^2}{f_{\rm PQ}^2} |H_u \cdot H_d|^2 \nonumber\\
&&+ \frac{g_{\rm PQ}^2}{2}\frac{m_C^2}{m_{V_{\rm PQ}}^2 + m_C^2} \left(\sum_a q_a|H_a|^2 \right)^2 + V_{\rm soft} \ ,
\end{eqnarray}
where $m_C^2 = 2(\sum_i m_{S_i}^2 f_i^2 q_i^2) / (\sum_i f_i^2 q_i^2)$, 
and $X=e^{A/f_{\rm PQ}}$. Here the auxiliary fields $F_A$, $F_A^\dagger$ and $D_{\rm PQ}$ have been integrated out. Note, the Wess-Zumino gauge did not fix the gauge transformation.
Selecting unitary gauge and integrating out the massive saxion mode, we get Eq.(\ref{Higgs_potential}) and Eq.(\ref{coefficients}), with $q_{H_u} = q_{H_d} = -\frac{1}{2}q_S $ assumed.

\subsection{A.2 Effective Higgs Potential: Super-unitary Gauge}

With unbroken SUSY, the PQ theory is invariant under the super-gauge transformation 
\begin{eqnarray}
{\bf A} \to {\bf A}+ f_{\rm PQ} {\bf \alpha} , \  {\bf V}_{\rm PQ} \to {\bf V}_{\rm PQ} - \frac{{\bf \alpha}+{\bf\alpha}^\dagger}{2g_{\rm PQ}} , \ {\bf H}_a \to {\bf H }_a e^{\bf \alpha}, \nonumber 
\end{eqnarray} 
where ${\bf \alpha}$ is a chiral superfield. In the case of super-unitary gauge, the K\"{a}hler potential and superpotential in Eq.(\ref{10}) can be rewritten as
\begin{eqnarray}
{\bf K} &\sim&  2 g_{\rm PQ}^2 f_{\rm PQ}^2 {\bf V}_{\rm PQ}^2  \nonumber \\&&+ 2 g_{\rm PQ} {\bf V}_{\rm PQ} \sum_a q_a {\bf H}_a^\dagger \exp({\bf U_{\rm SM}}) {\bf H}_a  \ , \nonumber \\
{\bf W}_{\rm H} &=& \lambda f_S  {\bf H_u H_d} \ .
\end{eqnarray}
Integrating out $\bf V_{\rm PQ}$, we have 
\begin{eqnarray}
{\bf K} = -\frac{g_{\rm PQ}^{2}}{m_{V_{\rm PQ}}^{2}} \left[ \sum_{a} q_{a }{\bf H}_a^\dagger \exp({\bf U_{\rm SM}}) {\bf H}_a \right]^{2} \ ,
\end{eqnarray}
which is the same as the one in~\cite{ArkaniHamed:1998nu} and leads to 
\begin{eqnarray}
V_{\rm SU} &\sim& \frac{2 g_{\rm PQ}^2\mu^2}{m_{V_{\rm PQ}}^2} (q_{H_u} + q_{H_d})^2 |H_u\cdot H_d|^2 \\
&&  + \frac{2 g_{\rm PQ}^2 \mu^2}{m_{V_{\rm PQ}}^2}  \left(q_{H_u} |H_u|^2 + q_{H_d} |H_d|^2\right)  \nonumber \\ && \times \left( q_{H_u} |H_d|^2 + q_{H_d} |H_u|^2 \right)  
\nonumber \\&& 
 - \frac{g_{\rm PQ}^2 (g_Y^2 + g_2^2)}{2 m_{V_{\rm PQ}}^2} \left(q_{H_u}|H_u|^4 + q_{H_d}|H_d|^4 \right. \nonumber \\&& \left.- (q_{H_u}+q_{H_d})|H_u|^2 |H_d|^2 \right) \left(\sum_a q_a|H_a|^2 \right) \ . \nonumber
 \end{eqnarray}
Although the Higgs fields have the power of six, the last term is comparable with the other ones, given that the Higgs VEVs and the $\mu$ parameter are of the same order.

The effects of the softly SUSY breaking can be incorporated via the interaction between the SUSY-breaking spurions and the superfields in the visible sector, which leads to new terms in {\bf K} and ${\bf W}_{\rm H}$  
\begin{eqnarray}
{\bf K}_{\rm PQ} &\sim& (- \sum_i m_{S_i}^2 \theta^2\bar\theta^2)f_i^2 e^{2q_i g_{\rm PQ} \bf V_{\rm PQ}}  \ ,\nonumber \\
{\bf W}_{\rm H} &\sim& (- A_\lambda \lambda \theta^2) f_S  \bf H_{u} \bf H_{d} \ ,
\end{eqnarray}
where $m_{S_i}^2$ is the soft squared mass of $S_i$ and $A_\lambda \lambda$ gives the $A$-parameter of $S H_u H_d$. $\bf V_{\rm PQ}$ has a general form (the metric $(-,+,+,+)$ is assumed)
\begin{eqnarray}
\bf V_{\rm PQ}(x,\theta,\bar\theta) &=& C(x) + i\theta\chi(x) - i\bar\theta\bar\chi(x) \nonumber \\&& + \frac{i}{2}\theta^2 [M(x)+iN(x)] \nonumber \\&& - \frac{i}{2}\bar\theta^2(M(x)-iN(x))  -\theta\sigma^\mu\bar\theta v_\mu(x)  \nonumber\\
&& + i\theta^2\bar\theta[\bar\lambda(x)  +\frac{i}{2}\bar\sigma^\mu\partial_\mu\chi(x)] \nonumber \\&&- i\bar\theta^2\theta[\bar\lambda+\frac{i}{2}\sigma^\mu\partial_\mu\bar\chi(x)]\nonumber\\
&&+\frac{1}{2}\theta^2\bar\theta^2[D(x) + \frac{1}{2}\Box C(x)] \ ,
\end{eqnarray}
with 
\begin{eqnarray}
\left.\bf V_{\rm PQ}^2\right|_{\theta^2\bar\theta^2} &\sim& - \frac{1}{2}v^\mu v_\mu + \frac{1}{2}(M^2 + N^2)  \nonumber\\&& + \frac{1}{2} C\Box C + CD \ , \nonumber \\ 
\left. \bf V_{\rm PQ} H_a^\dagger H_a\right|_{\theta^2\bar\theta^2} &\sim& \frac{C}{m_{V_{\rm PQ}}} F_a^\dagger F_a    + \frac{i}{2}[M+iN] H_a F_a^\dagger  \nonumber \\&& - \frac{i}{2}[M-iN] H_a^\dagger F_a \nonumber \\ && + \frac{1}{2}H_a^\dagger H_a [D + \frac{\Box C}{2 m_{V_{\rm PQ}}}]  \ , \\ 
 \left.  C  \bf H_a^\dagger U_{\rm SM} H_a\right|_{\theta^2\bar\theta^2} &\sim& 
\frac{ C}{m_{V_{\rm PQ}}} H_a^\dagger (Y_a g_Y D_Y + T^3 g_2 D_2^3) H_a \ . \nonumber 
\end{eqnarray}
Here Wess-Zumino gauge is assumed for the SM gauge superfields. Then we can get the effective Lagrangian
\begin{eqnarray}
{\cal L}_{\rm SU} &=& -\frac{1}{2} m_{V_{\rm PQ}}^2 v^\mu v_\mu + \frac{1}{2}m_{V_{\rm PQ}}^2 (M^2 + N^2) + m_{V_{\rm PQ}} C D  \nonumber\\
&&- \frac{1}{2} m_C^2 C^2 + 2g_{\rm PQ} \sum_a q_a \left( \frac{C}{m_{V_{\rm PQ}}} F_a^\dagger F_a  \right. \nonumber \\&& \left. + \frac{i}{2}[M+iN] H_a F_a^\dagger - \frac{i}{2}[M-iN] H_a^\dagger F_a  \right. \nonumber \\&& \left. + \frac{1}{2}H_a^\dagger H_a D   \right)  + \frac{1}{2}D^2 + \sum_a F_a^\dagger F_a \nonumber \\
&& + \lambda f_S  [H_{u} F_{d} +  F_{u} H_{d} + {\rm h.c.}] \nonumber \\&& + A_\lambda \lambda f_S [H_{u} H_{d} + {\rm h.c.}]  \\ 
&& + \frac{ 2 g_{\rm PQ} C}{m_{V_{\rm PQ}}}  \sum_a q_a H_a^\dagger (Y_a g_Y D_Y + T^3 g_2 D_2^3) H_a \ .  \nonumber 
\end{eqnarray}
Integrating out $M$, $N$, $D$, $C$, $D_Y$, $D_2^3$ and neglecting the mass term of $v_\mu$, 
we obtain the Higgs effective potential in super-unitary gauge
\begin{eqnarray}
V_{\rm SU} &=& (|\mu_{\rm eff}|^2 +  m_{H_u}^2)|H_u|^2 + (|\mu_{\rm eff}|^2 +  m_{H_d}^2)|H_d|^2 \nonumber\\
&& + \frac{1}{8}(g_2^2 + g_Y^2)(|H_u|^2-|H_d|^2)^2 - 2 B_\mu {\rm Re} (H_u H_d) \nonumber\\
&&+ \frac{2g_{\rm PQ}^2 \mu^2 (q_{H_u} + q_{H_d})^2}{m_{V_{\rm PQ}}^2}|H_u H_d|^2 \nonumber \\ &&
+ \frac{2 g_{\rm PQ}^2 \mu^2}{m_{V_{\rm PQ}}^2+m_C^2}\left(\sum_a q_a|H_a|^2 \right) \nonumber \\&& \times \left( q_{H_u} |H_d|^2 + q_{H_d} |H_u|^2 \right) \nonumber\\
&& + \frac{g_{\rm PQ}^2}{2}\frac{m_C^2}{m_{V_{\rm PQ}}^2 + m_C^2} \left(\sum_a q_a|H_a|^2 \right)^2 \nonumber \\ &&
- \frac{2g_{\rm PQ}^2}{m_{V_{\rm PQ}}^2+m_C^2}\left(\sum_a q_a|H_a|^2 \right)\left( \sum_a  m_{H_a}^2 q_a |H_a|^2 \right)  \nonumber \\
&& - \frac{g_{\rm PQ}^2 (g_Y^2 + g_2^2)}{2 (m_{V_{\rm PQ}}^2+ m_C^2)} \left(q_{H_u}|H_u|^4 + q_{H_d}|H_d|^4 \right. \nonumber \\&& \left.- (q_{H_u}+q_{H_d})|H_u|^2 |H_d|^2 \right) \left(\sum_a q_a|H_a|^2 \right) \ . 
\end{eqnarray}
So, the $A_\lambda$-term does not contribute the Higgs potential apart from giving a $B\mu$ term. 

\subsection{A.3 Scattering Amplitudes Of Light Fields}

Although the effective potential of the neutral Higgs fields is gauge-dependent, physical observables should not depend on the gauge option. Next, we show that the scattering amplitudes of the Higgs fields in the Wess-Zumino and super-unitary gauges are the same at tree level. For simplicity we will work in the limit of unbroken SUSY. In addition, no cubic terms appear in the tree-level effective potentials? 
Define ${\bf X}=e^{{\bf A}/f_{\rm PQ}}$, we have ${\bf X}^\dagger  {\bf X} = 1 +  {\bf Y}$ or $ {\bf Y}= \frac{{\bf A}+  {\bf A}^\dagger}{f_{\rm PQ}}$ in the Wess-Zumino gauge, with the decomposition 
\begin{eqnarray}
{\bf Y} \sim Y + \frac{1}{4}\theta^2\bar\theta^2 \Box Y \ ,
\end{eqnarray} 
where we have omitted the terms containing fermion fields and auxiliary fields. Then
${\bf K}_{\rm PQ} \sim \frac{1}{2}f_{\rm PQ}^2 {\bf Y}^2$
gives the kinetic term of $Y$ 
\begin{eqnarray}
{\cal L}_{\rm WZ}^{\rm k} \sim  \frac{1}{4} f_{\rm PQ}^2 Y\Box Y \ .
\end{eqnarray} 
Integrating out $Y$, we obtain a correction of the order $\mathcal O(\Box/m_{V_{\rm PQ}}^2)$
\begin{eqnarray}
{\cal L}_{\rm WZ}^{\rm k} &=& \frac{1}{4g_{\rm PQ}^4 f_{\rm PQ}^6} [ g_{\rm PQ}^2 f_{\rm PQ}^2(q_{H_u} |H_u|^2 + q_{H_d} |H_d|^2)\nonumber\\
&& +\lambda^2f_S^2 q_S (|H_u|^2 + |H_d|^2) ] \nonumber\\
&&\times \Box [ g_{\rm PQ}^2 f_{\rm PQ}^2(q_{H_u} |H_u|^2 + q_{H_d} |H_d|^2)\nonumber\\
&&+\lambda^2 f_S^2 q_S (|H_u|^2 + |H_d|^2) ]   \nonumber  \\
& \approx &  \frac{1}{4 f_{\rm PQ}^2} (q_{H_u} |H_u|^2 + q_{H_d} |H_d|^2)    \nonumber \\&& \Box (q_{H_u} |H_u|^2 + q_{H_d} |H_d|^2) \label{301}
\end{eqnarray}
where the corrections proportional to $\lambda^2$ and higher orders are neglected. 

With the super-unitary gauge, the kinetic term of the saxion field arises from 
\begin{eqnarray}	
{\bf V_{\rm PQ}} {\bf H}_a^\dagger {\bf H}_a |_{\theta^2\bar\theta^2} &\sim& \frac{C}{m_{V_{\rm PQ}}} (F_a^\dagger F_a + \frac{1}{4} H_a^\dagger\Box H_a \nonumber \\ && 
+ \frac{1}{4}\Box H_a^\dagger H_a - \frac{1}{2}\partial_\mu H_a^\dagger \partial^\mu H_a )\nonumber \\
&& + \frac{i}{2}[M+iN] H_a F_a^\dagger - \frac{i}{2}[M-iN] H_a^\dagger F_a \nonumber \\ &&  + \frac{1}{2}H_a^\dagger H_a [D + \frac{\Box C}{2 m_{V_{\rm PQ}}}] \ .
\end{eqnarray}
The Langrangian is given by
\begin{eqnarray}
{\cal L}_{\rm SU}^{\rm k} 
&\sim&   \frac{g_{\rm PQ}^2}{2m_{V_{\rm PQ}}^2}\left(\sum_a q_a|H_a|^2\right) \Box \left(\sum_a q_a|H_a|^2\right) \nonumber \\
&& - \frac{g_{\rm PQ}^2}{m_{V_{\rm PQ}}^2} \left( \sum_a q_a |H_a|^2 \right) \sum_a q_a \left[ \frac{1}{2}\Box(H^\dagger_aH_a)  \right. \nonumber \\&& \left. + \frac{1}{2} H_a^\dagger\Box H_a + \frac{1}{2}\Box H_a^\dagger H_a - \partial_\mu H_a^\dagger \partial^\mu H_a  \right] \ . \label{410}
\end{eqnarray}

The terms in the first line of Eq.(\ref{410}) are the same as the ones in Eq.~(\ref{301}), while the left ones give the difference of the kinetic terms between the Wess-Zumino gauge and the super-unitary gauge
\begin{eqnarray}     \label{312}
\Delta {\cal L}_{\rm k} &=& - \frac{g_{\rm PQ}^2}{m_{V_{\rm PQ}}^2} \left( \sum_a q_a |H_a|^2 \right) \sum_a q_a \left[ \frac{1}{2}\Box(H^\dagger_aH_a) \right. \nonumber \\ && \left. +\frac{1}{2} H_a^\dagger\Box H_a + \frac{1}{2}\Box H_a^\dagger H_a - \partial_\mu H_a^\dagger \partial^\mu H_a  \right] \nonumber\\
&=& - \frac{g_{\rm PQ}^2}{m_{V_{\rm PQ}}^2} \left( \sum_a q_a |H_a|^2 \right)    \\    && \times  \sum_a q_a \left[ (\Box H_a^\dagger) H_a + H_a^\dagger (\Box H)\right]  \nonumber  \\
&\rightarrow& -\frac{2g_{\rm PQ}^2 }{m_{V_{\rm PQ}}^2} \left( \sum_a q_a |H_a|^2 \right) \left(\sum_a q_a \tilde m_{H_a}^2 |H_a|^2\right)  \ ,  \nonumber 
\end{eqnarray}
where the equation of motion of $H_a$ as a free field has been used since in a scattering process the incoming and outgoing states are on-shell free particles. If SUSY is conserved, $H_u$ and $H_d$ can only get masses from the superpotential and we have $\tilde m_{H_a} = \lambda f_S$. Then, with $m_{V_{\rm PQ}}^2 = 2 g_{\rm PQ}^2 f_{\rm PQ}^2$, it is easy to check that the potential difference is completely compensated by Eq.(\ref{312}), and therefore at tree-level, the scattering amplitudes of $H_{u,d}$ in the Wess-Zumino and super-unitary gauges are equivalent to each other. 

\subsection{A.4 Vacuum Energy And Particle Mass Spectrum}

The potentials in the Wess-Zumino and super-unitary gauges can be written as
\begin{eqnarray}
V_{\rm WZ} = V^{(0)} +  V^{(1)}_{\rm WZ} \ , \;\; V_{\rm SU} = V^{(0)} + V^{(1)}_{\rm UG} \ ,
\end{eqnarray}
with the minimization conditions given by
\begin{eqnarray}
\left.\frac{\partial V_{\rm WZ}}{\partial H_{u,d}} \right|_{H_u = v_u^{\rm WZ}, H_d = v_d^{\rm WZ}} &=& 0 \ , \nonumber \\
\left.\frac{\partial V_{\rm SU}}{\partial H_{u,d}} \right|_{H_u = v_u^{\rm SU}, H_d = v_d^{\rm SU}} &=& 0 \ . 
\end{eqnarray}
Here $V^{(1)}_{WZ}$ and $V^{(1)}_{UG}$ are of order $\sim \mathcal O(\lambda^2)$ ($m^2_{\rm soft}/m_{V_{\rm PQ}}^2 \sim \lambda^2$ has been assumed). Similarly, the VEVs in the two gauges can be written as
\begin{eqnarray}
v_{u,d}^{\rm WZ} = v_{u,d}^{(0)} +v_{u,d}^{\rm WZ(1)}\ ,\;\; v_{u,d}^{\rm SU} = v_{u,d}^{(0)} +  v_{u,d}^{\rm UG(1)}\ ,
\end{eqnarray}
with $v_{u,d}^{(0)}$ satisfying the minimization conditions of $V^{(0)}=V_{\rm MSSM}$
\begin{eqnarray}
m_{H_u}^2 + |\mu|^2 - B_\mu  \cot \beta -\frac{m_Z^2}{2} \cos 2\beta &=& 0, \nonumber \\ 
m_{H_d}^2 + |\mu|^2 - B_\mu  \tan \beta +\frac{m_Z^2}{2} \cos 2\beta &=& 0.  \label{400}
\end{eqnarray}
Then, up to $\sim \mathcal O(\lambda^2)$ the minimization conditions are given by
\begin{eqnarray}
&&\left.\frac{\partial^{2} V^{(0)}}{\partial H_{a} \partial H_{b}}\right|_{v_{u,d}^{(0)}}  v^{\rm WZ(1)}_{b} +  \left.\frac{\partial V_{\rm WZ}^{(1)}}{\partial H_{a}}\right|_{v_{u,d}^{(0)}} = 0 \ , \nonumber \\
&&\left.\frac{\partial^{2} V^{(0)}}{\partial H_{a} \partial H_{b}}\right|_{v_{u,d}^{(0)}}  v^{\rm UG(1)}_{b} +  \left.\frac{\partial V_{\rm SU}^{(1)}}{\partial H_{a}}\right|_{v_{u,d}^{(0)}} = 0 \ .  \label{403}
\end{eqnarray}
The difference of the potentials of the Wess-Zumino and super-unitary gauges is 
\begin{eqnarray}
\Delta V &=&  V_{\rm WZ} - V_{\rm SU}   \nonumber \\ 
&=&\frac{1}{f_{\rm PQ}^2} (|H_u|^2 + |H_d|^2) \nonumber \\&&\times  \left[V^{(0)} + \frac{1}{8} (g_2^2 + g_1^2)(|H_u|^2-|H_d|^2)^2 \right].
\end{eqnarray}
It is easy to check that
\begin{eqnarray}
\Delta V |_{v^{(0)}_{u,d}} \equiv 0 \ .
\end{eqnarray}
This indicates that the vacuum energies are equal to each other in the two different gauges, since
\begin{eqnarray}
(V^{(0)}+\Delta V)_{v^{(0)} + \Delta v} &=& V^{(0)}|_{v^{(0)}} + \left.\frac{\partial V^{(0)}}{\partial H} \right|_{v^{(0)}} v^{(1)} \nonumber \\ && + \Delta V |_{v^{(0)}} + {\rm higher~order}.
\end{eqnarray}
and the first three terms are zero.

Next, let us check whether the pole mass of the SM gauge bosons and fermions is the same in these two gauges. 
In the Wess-Zumino gauge the mass formulae of these particles are canonical, while they are not in the super-unitary gauge. In the latter case, the formulae are  
\begin{eqnarray}
m^2_{\rm W} &=&\left(1+\frac{2 g_{\rm PQ} C}{m_{V_{\rm PQ}}}\right) \frac{g_2^2}{2} v_{\rm EW}^2 \ , \nonumber  \\ 
m^2_{\rm Z} &=&\left(1+\frac{2 g_{\rm PQ} C}{m_{V_{\rm PQ}}}\right)   \frac{g_2^2+g_Y^2}{2} v_{\rm EW}^2 \ , \nonumber  \\ 
m_t&=& \left( 1 + \frac{g_{\rm PQ}  C}{m_{V_{\rm PQ}}} \right) \frac{y_t}{\sqrt 2} v_u t_L t_R \ , \label{401}
\end{eqnarray}
with 
\begin{eqnarray}
C  = - \frac{g_{\rm PQ}}{m_{V_{\rm PQ}}}(q_{H_u}|H_u|^2 + q_{H_u}|H_d|^2) \  .
\end{eqnarray}
Here we select top quark as an example of the SM fermions. 

From Eq.~(\ref{403}) we get
\begin{eqnarray}
v_b^{(1)} = - \left.(M^{2}_H)^{-1}_{ba} \frac{\partial \Delta V}{\partial H_a} \right|_{v^{(0)}_{u,d}} \ ,
\end{eqnarray}
where the entries $(M^2_H)_{ab} = \left.\frac{\partial V^{(0)}}{\partial H_a \partial H_b}\right|_{v^{(0)}_{u,d}}$ are
\begin{eqnarray}
(M^2_H)_{11} &=& 2 B_\mu \cot\beta + 2{m^{(0)}_Z}^2 \sin^2\beta  \nonumber \\
(M^2_H)_{12}&=&  -2B_\mu - 2{m^{(0)}_Z}^2 \sin\beta\cos\beta \nonumber \\
(M^2_H)_{21}&=&   -2B_\mu - 2{m^{(0)}_Z}^2 \sin\beta\cos\beta \nonumber \\
(M^2_H)_{22}&=&  2B_\mu \tan\beta + 2{m^{(0)}_Z}^2 \cos^2\beta  \ ,
\end{eqnarray}
and $\left.\frac{\partial \Delta V}{\partial H_a} \right|_{v^{(0)}_{u,d}}$ are
\begin{eqnarray}
\left.\frac{\partial \Delta V}{\partial H_u}\right|_{v_{u,d}^{(0)}}  = -\frac{{m^{(0)}_Z}^2 \cos2\beta v^3 \sin\beta}{f_{\rm PQ}^2} \ , \nonumber \\
\left.\frac{\partial \Delta V}{\partial H_d}\right|_{v_{u,d}^{(0)}} =  \frac{{m^{(0)}_Z}^2 \cos2\beta v^3 \cos\beta}{f_{\rm PQ}^2}\ . \nonumber \\
\end{eqnarray}
with $v^2 \equiv (H_u^2 + H_d^2 )_{v^{(0)}}$. This leads to
\begin{eqnarray}
 v_u^{(1)} = \frac{v^2 v_u^{(0)}}{2f_{\rm PQ}^2} \ ,\;\; v_d^{(1)} = \frac{v^2 v_d^{(0)}}{2f_{\rm PQ}^2} \ .
\end{eqnarray}
Therefore, the Higgs VEVs are rescaled by a factor $\frac{v^2}{2 f_{\rm PQ}^2}$ which cancels the  
rescaling factors in Eq.(\ref{401}) exactly. 

\section{B Accidental cancelation in the light scalar exotic scenario}

\begin{figure}[ht]
\begin{center}
\includegraphics[width=0.35\textwidth]{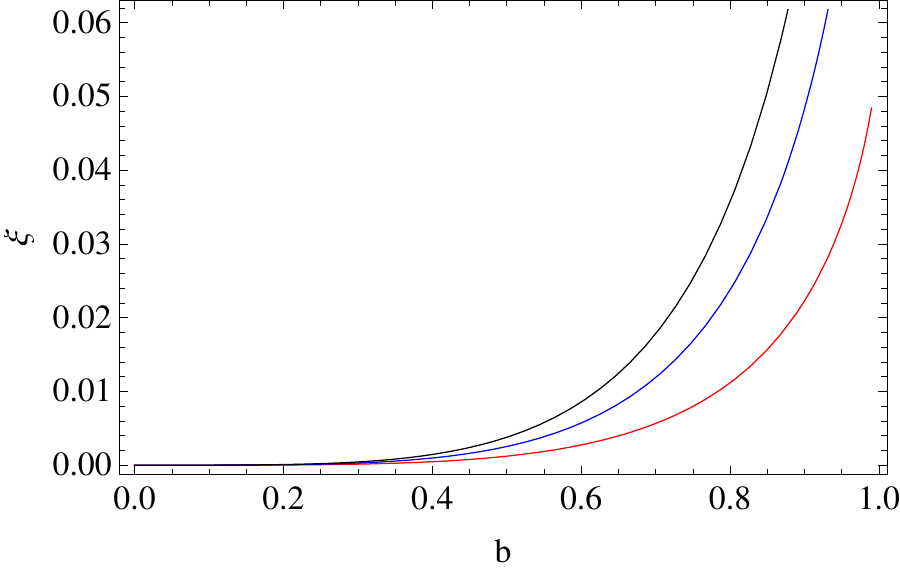}
\caption{$\xi$ vs. $b$ for varied $a$ values, where the red, blue and black curves are for $a$=1, 1.5 and 2, respectively.}
\label{xi}
\end{center}
\end{figure}

In the case of light charged scalar spectators coupled to Higgs fields through large A-terms, there is an accidental cancelation in the calculation of $\Delta T$. To see this point, let us discuss a simpler model, which is the stau-like particles, with their mass matrix scaled to  
$
{\cal M}^2_{\tilde\tau} = m_0^2 \begin{pmatrix}
1 & ab \\
  ab & a^2 
\end{pmatrix}
$. 
Here $a$ is the soft mass parameter of the right-chiral stau and $b$ is the mixing parameter. In the limit of $b=1$ where the mixing between the left- and right-chiral stau leptons and hence the weak isospin violation of the left-chiral stau doublet are maximized, $\Delta T$ is proportional to a factor
\begin{eqnarray}
 \xi = \log (1 + a^2) -  \frac{2}{1 + a^2} \int_0^1 dy (a^2 y + 1) \log(a^2 y + 1).  \nonumber
\end{eqnarray}
An interesting observation is that the numerical values of the two terms in the r.h.s are accidentally close to each other, which leads to a $\Delta T$ a few times smaller than its natural value. This effect can be generalized to varied $b$ values (see Fig.~\ref{xi}). $\tan\beta\sim 1$ becomes unnecessary therefore to avoid a sizable $\Delta T$.

\vspace*{-15pt}


\end{document}